\documentstyle[preprint,aps]{revtex}

\begin{document}

\title{Electromagnetic interaction between two uniformly moving 
charged particles: \\ a geometrical derivation using Minkowski 
diagrams}

\author{C\u alin Galeriu}

\address{Physics Department, Clark University, Worcester, MA, 01610, USA}

\maketitle

\begin{abstract}
This paper presents an intuitive, geometrical derivation of 
the relativistic addition of velocities, and of the 
electromagnetic interaction between two uniformly moving charged 
particles, based on 2 spatial + 1 temporal dimensional Minkowski 
diagrams. We calculate the relativistic addition of velocities by projecting 
the world-line of the particle on the spatio-temporal planes of the 
reference frames considered. We calculate the real component of the electromagnetic 
4-force, in the proper reference frame of the source particle, starting 
from the Coulomb force generated by a charged particle at rest. We then 
obtain the imaginary component of the 4-force, in the same reference 
frame, from the requirement that the 4-force be orthogonal to the 
4-velocity. The 4-force is then projected on a real 3 dimensional 
space to get the Lorentz force.
\end{abstract}
   
\section{Introduction}
Special Relativity, as presented in today's textbooks, is a complex 
mathematical theory. The 1 spatial + 1 temporal dimensional Minkowski 
diagrams , which initially introduce the Lorentz transformation, the 
time dilatation and the length contraction, are soon put aside in favor 
of an approach based on differential calculus and linear algebra. One 
gets little intuitive understanding of the law of relativistic addition 
of velocities, and of the fact that "magnetism is a kind of 
'second-order' effect arising from relativistic changes in the electric 
fields of moving charges" \cite{purcell}. However, by introducing only slightly more elaborate Minkowski diagrams, and using geometrical 
derivations, one can get back the intuitive understanding, to the 
great delight of the physicist who still believes in the spirit of 
Descartes' philosophy. 

\section{Real plane and complex plane: same trigonometry}
The complex plane, like the real plane, is a two-dimensional (2D) 
vector space. The scalar products for the real plane (1) and for 
the complex plane (2) are defined as follows
\begin{equation}
\hat \textbf{x} \cdot \hat \textbf{x} = 1 \hskip.5in \hat \textbf{y} 
\cdot \hat \textbf{y} = 1 \hskip.5in \hat \textbf{x} \cdot \hat \textbf{y} = 
\hat \textbf{y} \cdot \hat \textbf{x} = 0 
\end{equation}
\begin{equation}
\hat \textbf{x} \cdot \hat \textbf{x} = 1 \hskip.5in \hat \textbf{i} \cdot \hat 
\textbf{i} = -1 \hskip.5in \hat \textbf{x} \cdot \hat \textbf{i} = \hat \textbf{i} 
\cdot \hat \textbf{x} = 0 
\end{equation}
where $\hat \textbf{x}$,$\hat \textbf{y}$ and respectively $\hat \textbf{x}$,$\hat \textbf{i}$ are the basis vectors.

Since both planes have a scalar product, one can talk about orthogonal 
vectors (their scalar product is zero) and about the magnitude of a 
vector (the square root of the scalar product of a vector with 
itself). This allows us to define the circle (the geometric locus 
of the points equally spaced from a given point), the angle in 
radians (the length, between two points of a circle of radius one, 
measured along the circumference), and the trigonometric functions 
sine and cosine (the magnitudes of the projections of a radius one 
vector on the two coordinate axes). From these definitions it follows 
that, for both the real and the complex planes, one has the relations:
\begin{equation}
\sin^2(\alpha) + \cos^2(\alpha) = 1  
\end{equation}
\begin{equation}
[ {d \over d\alpha} \sin(\alpha) ]^2 + [ {d \over d\alpha} \cos(\alpha) ]^2 = 1.
\end{equation}
From (3)-(4) one can get the derivatives of the trigonometric functions, 
the Taylor series expansions of sine and cosine, and then all the 
well known trigonometric relations. The only detail we have to keep 
in mind is that the angle $\alpha$ in the real plane is a real 
number, while in the complex plane it is a purely imaginary number, 
due to the non-positive definite scalar product used in the last case. There 
are a few more relevant differences, which we can best point out if 
we represent the complex plane as an Euclidean plane. Two vectors in 
the complex plane are orthogonal if they make the same angle with the 
bisecting line of the first quadrant. The circle in the complex 
plane looks like a hyperbola \cite{minkowski}. Not any line passing through the 
origin intersects the right (or left) branch of the hyperbola. This means 
that there are pairs of lines passing through the origin to which 
we cannot assign an angle. However, for the triangles we will be 
working with, the ratio of segments behaves as if it were an angle, 
of negative value. The true angle is obtained by symmetry with 
respect to the first bisecting line, as the pair $\alpha$ and 
$-\alpha$ indicates in Figure~\ref{velocity}.

\section{Relativistic addition of velocities}
Consider a reference frame K' which is moving with a velocity 
$\textbf{V} = V \hat\textbf{x}$ relative to another one K, and a particle 
moving with a velocity $\textbf{v}' = v'_x \hat\textbf{x}' + v'_y \hat\textbf{y}' 
+ v'_z \hat\textbf{z}'$ in the reference frame K'. The reference frames 
are chosen such that their origins and the particle coincide at the 
space-time point O, as shown in Figure~\ref{velocity}. Notice that $\hat\textbf{y} = 
\hat\textbf{y}'$ and $\hat\textbf{z} = \hat\textbf{z}'$, because $\textbf{V}$ has a 
component only in the x direction. The Oz axis is not plotted, but is similar to the Oy axis. The question is: What is the 
velocity $\textbf{v} = v_x \hat\textbf{x} + v_y \hat\textbf{y} + v_z \hat\textbf{z}$ 
of the particle in the reference frame K?

The world-line OP of the particle is projected on the complex planes 
(x,O,ict), (y,O,ict), (z,O,ict), (x',O,ict'), (y',O,ict'), 
(z',O,ict'), and the resulting angles from the respective projections 
give the components of the velocity of the particle in the two 
reference frames considered. For the situation considered the planes 
(x,O,ict) and (x',O,ict') coincide. It is seen from Figure~\ref{velocity} that
\begin{equation}
\tan (-\alpha) = {EF \over OE} = {V \over i c} 
\hskip.75in \tan (\alpha) = -{V \over i c} 
\end{equation}
\begin{equation}
\tan (-\beta) = {DC \over OD} = {v'_x \over i c} 
\hskip.75in \tan (\beta) = - {v'_x \over i c} 
\end{equation}
\begin{equation}
\tan (-\gamma) = {DA \over OD} = {v'_y \over i c} 
\hskip.75in \tan (\gamma) = -{v'_y \over i c} 
\end{equation}
\begin{equation}
\tan (-\delta) = {EB \over OE} = {v_y \over i c} 
\hskip.75in \tan (\delta) = -{v_y \over i c} 
\end{equation}
\begin{equation}
\tan (-\theta) = {EC \over OE} = {v_x \over i c} 
\hskip.75in \tan (\theta) = -{v_x \over i c}.
\end{equation}
In order to express $v_x$ and $v_y$ as functions of $V,v'_x$ and 
$v'_y$ we need to express $\delta$ and $\theta$ as functions of 
$\alpha$,$\beta$ and $\gamma$. 

In the plane (x,O,ict) of the Lorentz boost the addition of 
velocities is based on the addition of angles \cite{einstein}
\begin{equation}
\theta = \alpha + \beta 
\end{equation}
\begin{equation}
\tan (\theta) = \tan (\alpha + \beta) = {\tan (\alpha) + \tan (\beta) 
\over 1 - \tan (\alpha) \tan (\beta)}.
\end{equation}
From (11), by substitution of the tangents (5)-(9), it follows that
\begin{equation}
v_x = {V + v'_x \over 1 + Vv'_x/c^2}.
\end{equation}
Two rectangles, APCD and BPCE, result from the projection process. 
It is evident that
\begin{equation}
{CP \over OC} = {EB \over OC} = {EB \over OE}{OE \over OC} = \tan 
(-\delta) \cos (-\theta)
\end{equation}
\begin{equation}
{CP \over OC} = {DA \over OC} = {DA \over OD}{OD \over OC} = \tan 
(-\gamma) \cos (-\beta).
\end{equation}
From (13)-(14) it follows that
\begin{equation}
\tan (\delta) = {\cos (\beta) \tan (\gamma) \over \cos 
(\alpha + \beta)} = {\tan (\gamma) \over \cos (\alpha) 
[1 - \tan (\alpha) \tan (\beta)]}.
\end{equation}
By substitutions of the tangents (5)-(8) and of $\cos (\alpha) = 
[1 + \tan ^2(\alpha)]^{-1/2}$ we get
\begin{equation}
v_y = {v'_y (1 - V^2/c^2)^{1/2} \over 1 + Vv'_x/c^2}.
\end{equation}
A similar expression is obtained for the $v_z$ component.

\section{Electromagnetic interaction between two uniformly moving 
charged particles}
Consider two charged particles (with charges $Q_1$ and $Q_2$) at 
some arbitrary positions, moving with arbitrary, but uniform, 
velocities. We orient our 3D reference frame in such a way that 
the first particle (which generates the field) is initially at 
the origin,  moving along the Ox axis with velocity 
$\textbf{V} = V \hat\textbf{x}$, and the vector $\textbf{R} = R \cos (\theta) 
\hat\textbf{x} + R \sin (\theta) \hat\textbf{y}$ connecting the two particles 
is in the (x,O,y) plane. The angle between $\textbf{R}$ and the Ox axis 
is $\theta$. The second particle (subject to the electromagnetic 
field generated by the first one) is moving with velocity 
$\textbf{v} = v_x \hat\textbf{x} + v_y \hat\textbf{y} + v_z \hat\textbf{z}$. A section 
through the (x,O,y) plane can be seen in Figure~\ref{eminter}. The first particle 
is at point O and the second one is at point A. 

\subsection{Analytical calculation of the Lorentz force}
The electric field (in Gaussian units) generated by the first 
particle at the position of the second particle is \cite{landau,donnelly}
\begin{equation}
\textbf{E} = {Q_1 \textbf{R} \over R^3} (1 - {V^2 \over c^2}) 
[1 - {V^2 \over c^2} \sin ^2 (\theta)]^{-3/2}.
\end{equation}
The magnetic field generated by the first particle is
\begin{equation}
\textbf{H} = {1 \over c} \textbf{V} \times \textbf{E}.
\end{equation}
The Lorentz force acting on the second particle is
\begin{equation}
\textbf{F} = Q_2 \textbf{E} + {Q_2 \over c} \textbf{v} \times \textbf{H}.
\end{equation}
From (17)-(19) the Cartesian components of the force \cite{rosser} are obtained
\begin{equation}
F_x = {Q_1 Q_2 \over R^2} (1 - {V^2 \over c^2}) 
[1 - {V^2 \over c^2} \sin ^2 (\theta)]^{-3/2} 
[\cos (\theta) + \sin (\theta) {v_y V \over c^2}] 
\end{equation}
\begin{equation}
F_y = {Q_1 Q_2 \over R^2} (1 - {V^2 \over c^2}) [1 - {V^2 \over c^2} 
\sin ^2 (\theta)]^{-3/2}  \sin (\theta) (1 - {v_x V \over c^2}) 
\end{equation}
\begin{equation}
F_z = 0.
\end{equation}

\subsection{Geometrical derivation of the Lorentz force}
The force components (20)-(22) can be obtained in a more graphical way, if 
we start with the Coulomb force generated by a charged particle at 
rest. One key assumption or experimental fact is that in a frame 
where all the source charges producing an electric field $\textbf{E}$ 
are at rest, the force on a charge $q$ is given by $\textbf{F} = q \textbf{E}$ 
independent of the velocity of the charge in that frame \cite{jackson}. The 
reference frame K' in which the source particle is at rest is moving 
with velocity $\textbf{V}$ relative to the original frame K.

In the reference frame K the particle at A is observed to interact 
with the particle at O. The distance between particles is $R$, the 
length of the segment $OA$.

In the reference frame K' the particle at A is observed to interact 
with the particle at B, where the segment $BA$ is a position vector 
$\textbf{R}'$ parallel to the plane (x',O,y'). The following construction
gives the position of point B: the segment $AE$ is 
parallel to Oy and intersects the Ox axis at E, whereas the segment 
$EB$ is parallel to Ox' and intersects the world-line $CO$ at B. 
$BD$ projects the point B on the Ox axis at D.

Relative to K', the particle at B exerts a radial Coulomb force on 
the particle at A. This force (in Gaussian units) is 
\begin{equation}
\textbf{F}' = {Q_1 Q_2 \over R'^3} \textbf{R}' 
\end{equation}
where $\textbf{R}' = R' [\cos (\theta ') \hat\textbf{x}' + \sin (\theta ') 
\hat\textbf{y}']$.

The key point in getting the force $\textbf{F}$ in the reference frame K 
is to notice that the force, in any reference frame considered, is 
given by the projection on the real 3D space of that frame of the 
4-force $\bbox{\cal F}$ (which is a Minkowski-space vector), that is
\begin{equation}
\bbox{\cal F} = \bbox{\cal F}_{\rm real} + \bbox{\cal F}_{\rm imag} 
= \gamma(v) \textbf{F} + \hat\textbf{i} \gamma(v) {P \over c}
\end{equation}
\begin{equation}
\bbox{\cal F} = \bbox{\cal F}'_{\rm real} + \bbox{\cal F}'_{\rm imag} 
= \gamma(v') \textbf{F}' + \hat\textbf{i}' \gamma(v') {P' \over c}
\end{equation}
where $\gamma(v) = (1 - v^2/c^2)^{-1/2}$ and $P = \textbf{F}\cdot\textbf{v}$.

We will obtain the 4-force $\bbox{\cal F}$ from its real and imaginary 
components ($\bbox{\cal F}'_{\rm real}$ and $\bbox{\cal F}'_{\rm imag}$) in 
the reference frame K', then we will decompose the same 4-force into 
its real and imaginary components ($\bbox{\cal F}_{\rm real}$ and 
$\bbox{\cal F}_{\rm imag}$) in the reference frame K. The Lorentz force 
we are looking for is just $\textbf{F} = \bbox{\cal F}_{\rm real}/
\gamma(\textbf{v})$.

From (23)-(25) it follows that
\begin{equation}
\bbox{\cal F}'_{\rm real} = \gamma(v') {Q_1 Q_2 \over R'^3} \textbf{R}'.
\end{equation}
To get the imaginary component $\bbox{\cal F}'_{\rm imag}$ we use the 
orthogonality between the 4-force and the 4-velocity, 
$\bbox{\cal F}\cdot\bbox{\cal V} = 0$, where the 4-velocity is 
$\bbox{\cal V} = \gamma(v') \textbf{v}' + \hat\textbf{i}' \gamma(v') c$. 
The orthogonality condition leads to
\begin{equation}
\gamma^2(v') {Q_1 Q_2 \over R'^2} {\textbf{R}'\cdot\textbf{v}' \over R'} 
+ \bbox{\cal F}'_{\rm imag}\cdot\hat\textbf{i}' \gamma(v') c = 0 
\end{equation}
\begin{equation}
\bbox{\cal F}'_{\rm imag} = \hat\textbf{i}' \gamma(v') {Q_1 Q_2 \over R'^2} 
{v'_{\rm rad} \over c}
\end{equation}
where the radial component of the velocity is
\begin{equation}
v'_{\rm rad} = {\textbf{R}'\cdot\textbf{v}' \over R'} = v'_x \cos (\theta ') 
+ v'_y \sin (\theta ').
\end{equation}
The components of the force $\textbf{F}$ in the reference frame K are 
given by the projection of the 4-force $\bbox{\cal F}$ on the 3D real space 
of K. An easy way to do this is to notice that we can decompose  
$\bbox{\cal F}'_{\rm real}$ (which has the direction of the segment 
$BA$) and $\bbox{\cal F}'_{\rm imag}$ (which has the direction of the 
segment $BO$) into sums of 4-vectors, each of the 4-vectors being 
parallel to one of the axes of the reference frame K:
\begin{equation}
\textbf{r}_{BA} = \textbf{r}_{BD} + \textbf{r}_{DE} + \textbf{r}_{EA} 
\end{equation}
\begin{equation}
\textbf{r}_{BO} = \textbf{r}_{BD} + \textbf{r}_{DO}
\end{equation}
Because these expansions do not involve any component along the 
Oz axis, this simply means that $F_z = 0$. The projections of the 
4-force on the Ox and Oy axes are
\begin{equation}
\gamma(v) F_x = {\cal F}'_{\rm real} {DE \over BA} + {\cal F}'_{\rm imag} 
{DO \over BO}
\end{equation}
\begin{equation}
\gamma(v) F_y = {\cal F}'_{\rm real} {EA \over BA}.
\end{equation}
The lengths of the various segments needed above are as follows:
\begin{equation}
AO = R 
\end{equation}
\begin{equation}
EA = AO \sin (\theta) = R \sin (\theta) 
\end{equation}
\begin{equation}
OE = AO \cos (\theta) = R \cos (\theta) 
\end{equation}
\begin{equation}
BE = OE \cos (\alpha) = R \cos (\theta) \cos (\alpha) 
\end{equation}
\begin{equation}
DE = BE \cos (\alpha) = R \cos (\theta) \cos ^2(\alpha) 
\end{equation}
\begin{equation}
AB = (AE^2 + BE^2)^{1/2} = R \cos (\alpha) [1 + \tan ^2(\alpha) 
\sin ^2(\theta)]^{1/2} = R'
\end{equation}
We also notice that $DO/BO = \sin(-\alpha)$. The force components 
in (32)-(33) become
\begin{eqnarray}
F_x = &&{\gamma(v') \over \gamma(v)} {Q_1 Q_2 \over R^2} 
{\cos (\theta) \over \cos (\alpha) [1 + \tan ^2(\alpha) 
\sin ^2(\theta)]^{3/2}} \nonumber \\
&&+ i {\gamma(v') \over \gamma(v)} {Q_1 Q_2 \over R^2} {v'_{\rm rad} 
\over c} {\sin (-\alpha) \over \cos ^2(\alpha) [1 + \tan ^2(\alpha) 
\sin ^2(\theta)]}
\end{eqnarray}
\begin{equation} 
F_y = {\gamma(v') \over \gamma(v)} {Q_1 Q_2 \over R^2} 
{\sin (\theta) \over \cos ^3(\alpha) [1 + \tan ^2(\alpha) 
\sin ^2(\theta)]^{3/2}}.
\end{equation}
We can also calculate
\begin{equation}
\sin (\theta ') = {EA \over AB} = {\sin (\theta) \over \cos (\alpha) 
[1 + \tan ^2(\alpha) \sin ^2(\theta)]^{1/2}} 
\end{equation}
\begin{equation}
\cos (\theta ') = {BE \over AB} = {\cos (\theta) \over 
[1 + \tan ^2(\alpha) \sin ^2(\theta)]^{1/2}}.
\end{equation}
If the velocity of the particle at A has the components 
$v_x,v_y,v_z$, as measured in the reference frame K, and K is moving 
with the velocity $\textbf{V}' = -V \hat\textbf{x}'$ relative to K', then 
the particle will have the following 
components of the velocity (compare with equations (12) and (16)) in the reference frame K'
\begin{equation}
v'_x = {v_x - V \over 1 - Vv_x/c^2} \hskip.5in v'_y = 
{v_y (1 - V^2/c^2)^{1/2} \over 1 - Vv_x/c^2} \hskip.5in v'_z = 
{v_z (1 - V^2/c^2)^{1/2} \over 1 - Vv_x/c^2}.
\end{equation}
With these components we find that 
\begin{equation}
\gamma (v') = \gamma (v) {(1 - Vv_x/c^2) \over (1 - V^2/c^2)^{1/2}} 
\end{equation}
and the radial velocity (29) becomes
\begin{equation}
v'_{\rm rad} = {(v_x - V) \cos (\alpha) \cos (\theta) + 
v_y (1 - V^2/c^2)^{1/2} \sin (\theta) \over (1 - V v_x/c^2) 
\cos (\alpha) [1 + \tan ^2(\alpha) \sin ^2(\theta)]^{1/2}}.
\end{equation}
Substituting $\gamma (v')$ and $v'_{rad}$ in (40)-(41), and also using the fact that $\sin(\alpha) = i (V/c) \gamma(V)$, 
$\cos(\alpha) = \gamma(V)$ and $\tan(\alpha) = i V/c$, 
we finally obtain the components in (20)-(21).

\section{Conclusions}
We have presented a geometrical calculation of the 
relativistic addition of velocities, and of the electromagnetic interaction 
between two uniformly moving charged particles. The geometrical approach used 
is an elegant, more intuitive and alternative way of obtaining these 
important results of Special Relativity. We hope our work will usefully 
complement other pedagogical efforts \cite{mermin1,mermin2,saletan} centered on Minkowski space diagrams.

\begin{figure}
\caption{Relativistic addition of velocities. The world-line OP is 
projected on various spatio-temporal planes. OA is the projection 
on (y',O,ict'), OB is the projection on (y,O,ict) and OC is the 
projection on (x,O,ict). The planes (x,O,ict) and (x',O,ict') coincide.}
\label{velocity}
\end{figure}

\begin{figure}
\caption{Electromagnetic interaction between two uniformly moving 
charged particles. CO is the world-line of the source particle, 
and AG is the world-line of the test particle. In the proper 
reference frame of the source particle there is a Coulomb force 
directed along the BA radial direction.}
\label{eminter}
\end{figure}

\end{document}